\documentclass{tcibook}
\usepackage{fancyhea}
\usepackage{work}
\usepackage{bm}       %    enables bold math symbols  e.g.  \bm{\gamma}
\usepackage{graphicx}
\usepackage{hyperref}      % hypertext links %%ARXIV
\usepackage{multirow}
\usepackage{lineno}

\newcommand{\nc}{\newcommand}  

\def\beq{\begin{equation}}
\def\eeq#1{\label{#1}\end{equation}}
\def\eeqn{\end{equation}}

\newenvironment{Eqnarray}%
   {\arraycolsep 0.14em\begin{eqnarray}}{\end{eqnarray}}
\def\beqa{\begin{Eqnarray}}
\def\eeqa#1{\label{#1}\end{Eqnarray}}
\def\eeqan{\end{Eqnarray}}

\nc{\ra}{\rightarrow}  
\nc{\slsh}{\slash\hspace*{-0.22cm}}
\def\Re{{\cal R \mskip-4mu \lower.1ex \hbox{\it e}\,}}
\def\Im{{\cal I \mskip-5mu \lower.1ex \hbox{\it m}\,}}

\nc{\vev}[1]{ \left\langle {#1} \right\rangle }
\nc{\bra}[1]{ \langle {#1} | }
\nc{\ket}[1]{ | {#1} \rangle }
\nc{\fb}{\,{\rm fb}^{-1}}
\nc{\ev}{{\rm eV}}
\nc{\kev}{{\rm keV}}
\nc{\Mev}{{\rm MeV}}
\nc{\gev}{{\rm GeV}}
\nc{\tev}{{\rm TeV}}
\nc{\mev}{{\rm MeV}}

\def\del{\partial}
\def\Dslash{\not{\hbox{\kern-4pt $D$}}}
\def\dslash{\not{\hbox{\kern-2pt $\del$}}}
\def\pslash{\not{\hbox{\kern-2pt $p$}}}
\def\ETmiss{ \not{\hbox{\kern-4pt $E$}}_T }

\def\ee{e^+e^-}

\def\msb{{\bar{\ssstyle M \kern -1pt S}}}

\begin{document}
%\linenumbers

\def\bibname{References}
\bibliographystyle{plain}

\raggedbottom

\pagenumbering{roman}

\parindent=0pt
\parskip=8pt
\setlength{\evensidemargin}{0pt}
\setlength{\oddsidemargin}{0pt}
\setlength{\marginparsep}{0.0in}
\setlength{\marginparwidth}{0.0in}
\marginparpush=0pt

% The content begins here

\pagenumbering{arabic}

\renewcommand{\chapname}{chap:intro_}
\renewcommand{\chapterdir}{.}
\renewcommand{\arraystretch}{1.25}
\addtolength{\arraycolsep}{-3pt}

\thispagestyle{empty}
\begin{centering}
\vfill

{\Huge\bf Planning the Future of U.S. Particle Physics}

{\Large \bf Report of the 2013 Community Summer Study}

\vfill

{\Huge \bf Chapter 6:  Accelerator Capabilities} 

\vspace*{2.0cm}
{\Large \bf Convener: W. Barletta}

\vfill

{\large  Study Conveners: M. Bardeen, W. Barletta, L.~A.~T.~Bauerdick, R. Brock,
D.~Cronin-Hennessy, M.~Demarteau, M.~Dine, J.~L. Feng, M. Gilchriese,
S. Gottlieb, J.~L.~Hewett, R. Lipton, H.~Nicholson, M.~E. Peskin,
S. Ritz, I.~Shipsey, H. Weerts}\\
\vspace{1cm}

{\large Division of Particles and Fields Officers in 2013:
J.~L. Rosner (chair), 
I. Shipsey (chair-elect), 
N. Hadley (vice-chair),
P. Ramond (past chair)}\\
\vspace{1cm}

{\large Editorial Committee:
R.~H. Bernstein,
N. Graf,
P. McBride,
M.~E. Peskin,
J.~L. Rosner,
N.~Varelas,
K. Yurkewicz}

\vfill

\end{centering}
\pagenumbering{roman}

\newpage
\mbox{\null}

\vspace{3.0cm}

{\Large \bf Authors of Chapter 6:}

\vspace{2.0cm}
 {\bf W. A. Barletta},
M.~Bai, 
M.~Battaglia, 
O.~Br\"uning,
 J.~Byrd,
 R.~Ent,
 J.~Flanagan,
W.~Gai, 
 J.~Galambos, 
G.~Hoffstaetter, 
M.~Hogan, 
M.~Klute, 
S.~Nagaitsev,
 M.~Palmer, 
S.~Prestemon, 
 T.~Roser,
L.~Rossi,
V.~Shiltsev,
G.~Varner,
K.~Yokoya

 \tableofcontents

\newpage

\mbox{\null}
\newpage
\pagenumbering{arabic}

%%%%%%%%%%%%%%%%%%%%%%%%%%%%%%%%%%%%%%%%%%%%%%%%%%%
%%%%%%%%%%%%%%%%%%%%%%%%%%%%%%%%%%%%%%%%%%%%%%%%%%%
%%%     All of your files should be in a subdirectory.  Here the
%%%     subdirectory is called Magnetism  .   The title of your
%%%     report should be   wgreport.tex in that subdirectory.  Input
%%%     that file here
%%%%%%%%%%%%%%%%%%%%%%%%%%%%%%%%%%%%%%%%%%%%%%%%%%%%
%%%%%%%%%%%%%%%%%%%%%%%%%%%%%%%%%%%%%%%%%%%%%%%%%%%

%\input Magnetism/wgreport.tex 

\def\lunit{ cm$^{-2}$s$^{-1}$}

\setcounter{chapter}{5}

\chapter{Accelerator Capabilities} 
\label{chap:ceo}

\begin{center}\begin{boldmath}

%\hyphenpenalty 10000

\begin{center}

\begin{large} {\bf Convener:  W. A. Barletta} \end{large}

M.~Bai, 
M.~Battaglia, 
O.~Br\"uning,
 J.~Byrd,
 R.~Ent,
 J.~Flanagan,
W.~Gai, 
 J.~Galambos, 
G.~Hoffstaetter, 
M.~Hogan, 
M.~Klute, 
S.~Nagaitsev,
 M.~Palmer, 
S.~Prestemon, 
 T.~Roser,
L.~Rossi,
V.~Shiltsev,
G.~Varner,
K.~Yokoya

\end{center}

%\hyphenpenalty 1000

%Conveners are also listed separately in authorlist.tex

\end{boldmath}\end{center}

\section{Introduction}

\begin{figure}[tb]
\begin{center}
\includegraphics[width=0.8\textwidth]{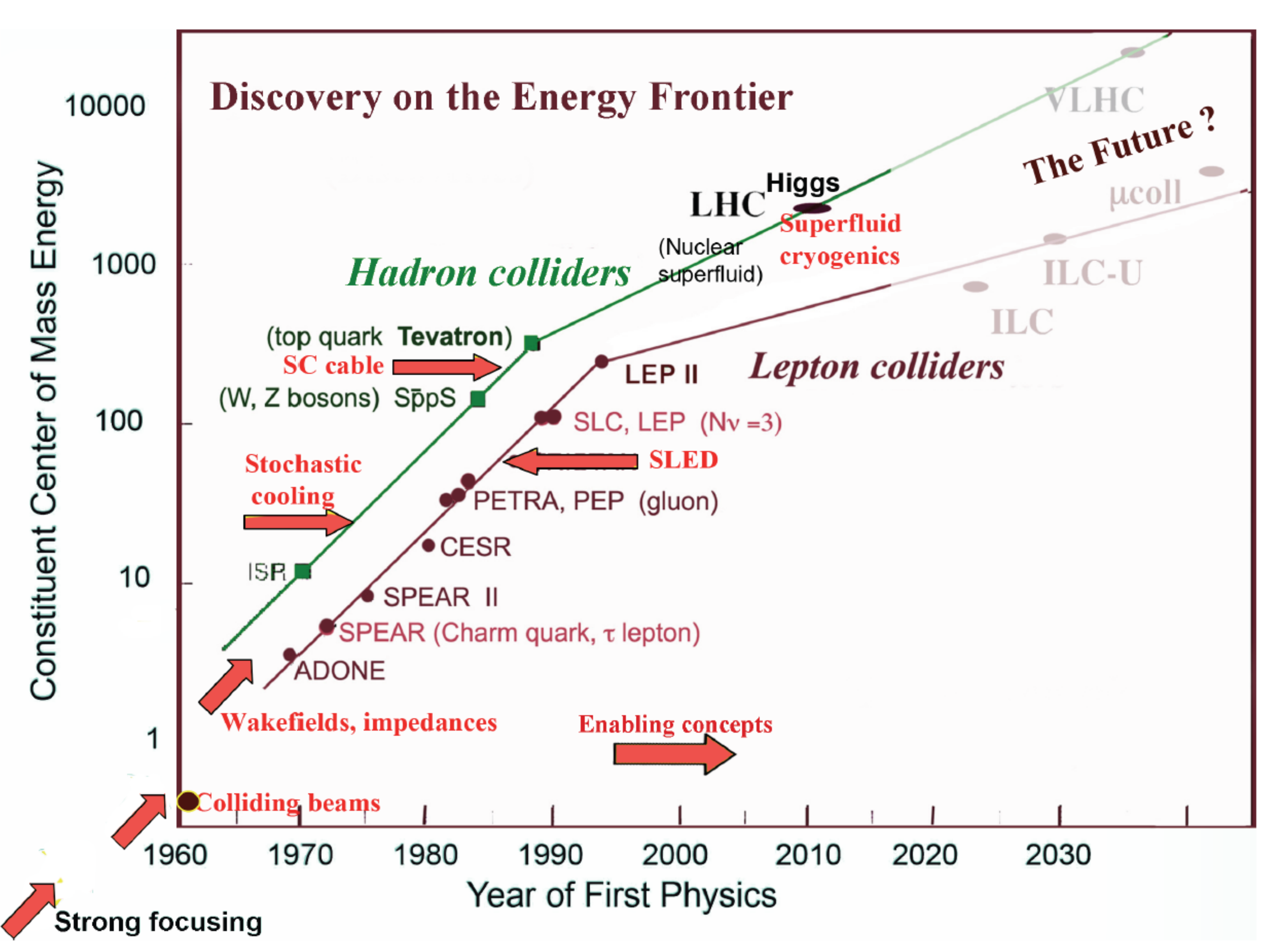}
\caption{The so-called Livingston plot illustrates how history of discovery on the energy frontier has been enabled by the history of invention (red arrows) in accelerator science and technology.}
\label{fig:Livingston}
\end{center}
\end{figure}

Accelerator-based experiments have been the mainstay and driving force of experimental particle physics ever since Rutherford recognized that radioactive sources provided an extremely limited capability to probe the nature of matter at the shortest time and distance scales.  From their earliest beginnings the history of invention in accelerator technology have enabled the history of particle physics discoveries especially on the energy frontier; see Fig.~\ref{fig:Livingston}.  

Yet the figure of merit of energy reach in the constituent center-of-momentum frame is 
far from the entire story of enabling capabilities. For experiments to be completed in a reasonable amount of time, the rate of interesting experimental events must be sufficiently high.  That is, the luminosity of the accelerator-target combination must be large enough to explore the physics of interest.
\beq
\mbox{Interesting\ Events} = \mbox{Cross\ section} \times \mbox{Average\ collision\ rate} \times \mbox{Time} \ . 
\eeq{Interesting}

%%%%%%%%%%%%%%%%%%%%%%%%%%%%%%%%%%%%
%%%%%%%%%%%%%%%%%%%%%%%%%%%%%%%%%

Generally high luminosity implies a large charge per bunch, a high collision frequency of beam bunches with the target, and focus of the beam to a small spot. The target can either be fixed or a counter-rotating beam bunch.
Even energy-reach and accelerator-target luminosity is insufficient for discovery-level experiments. The accelerator should deliver beam particles with appropriate quantum numbers and with a timing structure which maximizes the signal-to-background noise of the experiment. 

Assessing the suite of characteristics of present and proposed accelerators in the light of current interests and priorities of particle physicists has been the task of the Accelerator Capabilities Study of the Snowmass process. For proposed or planned capabilities the study has recommended the critical research and development programs needed.

The study has been a synthesis of individual pre-Snowmass workshops conducted by six working groups culminating with the collective Snowmass meeting of all interested participants. Each working group addressed major challenges foreseen for their respective class of accelerators in their respective pre-Snowmass workshops. The areas of inquiry included but were not limited to the following specific questions:

\begin{itemize}
\item   How high a luminosity is possible for the LHC? 
\item  	How high an energy is possible in the LHC tunnel? 
\item  	Could a Higgs factory be built in the LHC tunnel?
\item  	Can the ILC and  CLIC designs be improved using new technologies?  
\item  	Can one design a multi-TeV $\mu^+\mu^-$ collider?
\item  	What secondary beams are needed for Intensity Frontier experiments?
\item  	What proton beams are needed to generate these secondary beams, and can these be made by existing machines?
\item  	What accelerator capabilities at heavy flavor factories are required to realize the full range of physics opportunities? 
\item  	What are new physics opportunities using high-power electron and positron beams?
\item  	What is the broad range of test capabilities existing or needed for developing accelerator capabilities?
\end{itemize}

The working groups also considered a set of “big questions” regarding accelerator capabilities for the long-term future of high energy physics:
 \begin{itemize}
\item  	How can one build a collider at the 10 -- 30 TeV constituent mass scale?
\item  	What is the farthest practical energy reach of accelerator-based particle physics?
\item  	How would one generate ten or more megawatts of proton beam power? 
\item  	Can multi-megawatt targets survive and if so, for how long?
\item  	Can plasma-based accelerators achieve energies and luminosities relevant to particle physics?
\item  	Can accelerators be made an order of magnitude cheaper per GeV and/or per MW?
\end{itemize}

This report is organized as a collection of the executive summaries of each of the six working groups: Energy Frontier proton colliders, Energy Frontier lepton colliders, Intensity Frontier proton sources, Intensity Frontier electron accelerators, electron-ion colliders, and accelerator research and test-beam facilities.

\section{Energy Frontier proton colliders}

High-energy hadron colliders have been the tools for discovery at the highest mass scales of the Energy Frontier for more than a decade. They will remain so, unchallenged for the foreseeable future. The discovery of the Higgs boson at the LHC announced opened a new era for particle physics. Its measured properties are consistent, within the current uncertainties, with those of the Standard Model (SM) Higgs boson. 

After the discovery of the Higgs boson, understanding what is the real origin of electro-weak symmetry breaking becomes the next key challenge for collider physics. This challenge can be expressed in terms of two questions:  Up to what level of precision does the Higgs boson behave as predicted by the Standard Model? At what energy are the new particles that could offer some insight into the origin of dark matter, the matter-antimatter asymmetry, and neutrino masses? 

\subsection{The LHC and its upgrades}

The approved LHC program, its future upgrade towards higher luminosities (HL-LHC), and the study of an LHC energy upgrade or of a new proton collider delivering collisions at center-of-mass energies up to 100~TeV, are all essential components of this endeavor.  High luminosity could increase the precision of several key measurements, uncover rare processes, and guide and validate the progress in theoretical modeling, thus reducing systematic uncertainties in the interpretation of the data.  To achieve this end,

{\it the full exploitation of the LHC is the highest priority of the energy-frontier hadron-collider program. }

The LHC is expected to restart in Spring 2015 at a center-of-mass energy of 13--14~TeV and reach its  design luminosity of $10^{34}$cm$^{-2}$s$^{-1}$  during 2015.  After 2020, some critical components of the accelerator will reach the radiation damage limit, and others will have lower reliability due to vulnerability to radiation. Furthermore, as the statistical gain in running the accelerator without substantially increased luminosity will become marginal, the LHC will need a substantial increase in luminosity~\cite{ATLAS}. With a high luminosity upgrade (HL-LHC), the LHC should deliver 3000 fb$^{-1}$  during a decade of operation~\cite{Zimmermann}. The HL-LHC project is the first priority of Europe, as stated by the Strategy Update for High Energy Physics approved by CERN council in the special session of May 30, 2013 in Brussels.

Superconducting magnets are the single most critical technology for the LHC and for proton colliders beyond the present LHC level of perormance.  Indeed, the present LHC is based on 30 years of development of magnets using NbTi wire. In the LHC NbTi-based magnets are pushed to their limits both in the collider arcs and in the interaction regions designed and built in collaboration with U.S. and Japanese national laboratories. 

Delivering the beam brightness required for HL-LHC will be a difficult challenge for the upgraded LHC injector chain; also challenging will be preserving it in the LHC storage ring. In light of this limitation, the preferred route towards increased luminosity is reducing the beta (amplitude) function at the interaction point, 
$\beta^*$, by means of stronger triplet magnets with larger apertures in the interaction region. The present design of the HL-LHC interaction region requires quadrupoles with an aperture of 150~mm and with a peak field in excess of 12 T --- beyond the capabilities of NbTi conductor. It relies heavily on the success of the advanced Nb$_3$Sn technology developed by an integrated consortium of U.S. national laboratories under the LHC Accelerator Research Program (LARP). In addition to magnets, many other technologies will be involved in HL-LHC, such as crab cavities, advanced collimators, high-temperature superconducting links, and advanced remote handling. LARP plays a vital leadership role in this work to maximize the discovery potential of the extensive LHC infrastructure at CERN.

{\it A vigorous LARP leading to U.S. participation in the HL-LHC construction project is crucial to the full realization of the potential of LHC to deliver discovery physics to the thousands of U.S. researchers who are engaged at the LHC.}

Energy and luminosity play complementary roles in exploring the energy frontier with hadron collisions.  Production rates for signals of interest at the collider may be small either because the mass of the produced objects is large or because the coupling strength is small. In the former case, increasing the beam energy is clearly favored. In contrast, higher luminosity may be more effective in probing small couplings at smaller masses. In reaching higher energies, magnet technology will retain a pivotal role.

Given the progress in magnet technology and the maturity that Nb$_3$Sn is reaching, due largely to the LARP program for HL-LHC, Nb$_3$Sn magnets are expected to reach a limit of 15.5--16~T in operating field (that is with 15--20\% margin with respect to quench) 
within the next decade, thereby opening a path towards a 
collider with an energy significantly larger than the LHC. This would open the 
way towards very high energy hadron collisions with an exceptional potential for 
probing the Energy Frontier.
 
%%%%%%%%%%%%%%%%%%%%%%%%%%%%%%%%%%%%
\begin{figure}[tb]
\begin{center}
\includegraphics[width=0.95\textwidth]{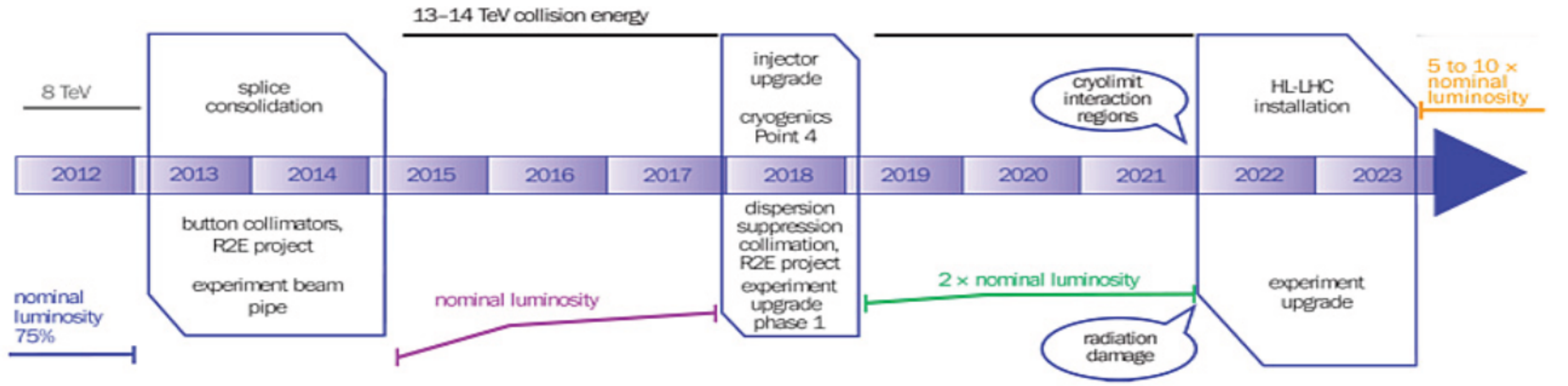}
\caption{LHC baseline plan for the next ten years in terms of energy of the collisions (upper line) 
and of luminosity (lower lines). The first long shutdown (LS) 2013-14 is to allow design parameters 
of beam energy and luminosity. The second LS, 2018, is to increase beam 
intensity and reliability as well as to upgrade the LHC Injectors.}
\label{fig:LHCsched}
\end{center}
\end{figure}
%%%%%%%%%%%%%%%%%%%%%%%%%%%%%%%%%

\subsection{Options for higher energy hadron colliders}

The first option is a machine housed in the LHC tunnel. The achievable center-of-mass energy depends on the dipole field strength. An energy of 26 TeV is within the reach of proven Nb$_3$Sn technology, although it requires engineering development. The energy reach would become 33 TeV, if 20 T magnets based on more futuristic high-temperature superconductors (HTS) were practical and affordable, and if they could fit in the limited space available in the LHC tunnel. In a new, longer tunnel, a  100~TeV proton collider is possible. Studies for a very large proton collider able to deliver center-of-mass energies of 100 to 200 TeV have been conducted over the past two decades. The multi-laboratory VLHC study led by Fermilab in 2001~\cite{VLHC}
was noteworthy. The analysis remains valid and argues for a very large (233 km) tunnel that allows the use of NbTi magnet technology.

Continuing an integrated, multi-laboratory program (LARP-like) towards the engineering development of 20~T magnets for HE-LHC or a new collider in a longer tunnel, needs to be an HEP priority. A program focused on engineering readiness, closely coordinated with CERN, would establish the limits of Nb$_3$Sn technology, investigate new conductor materials, and refine our present concepts for how to manage the enormous stresses produced by such high magnetic fields. The experience from RHIC, SSC, and LHC indicates that the dipoles account for about half of the total collider cost. Therefore, investing in magnet technology is critical and should represent a major focus of the research and development process.

With the renewed interest in the U.S. in a 100~TeV scale collider, the study group recommends participation in the international study for colliders in a large tunnel which CERN is now organizing.  The construction of a new tunnel reduces the demand for high dipole field strength, since there is a tradeoff between the tunnel circumference and required field strength (see Figure 3). The target collision energy is 100 TeV for 20 T dipoles in an 80 km tunnel ($R=13$ km). However, a slightly larger, 100 km circumference tunnel would provide the same collision energy of 100 TeV with field of 16 T, reachable with Nb$_3$Sn technology that relies on a far more mature and less expensive conductor than HTS. A large tunnel would also open up the possibility of a Higgs Factory based upon an $e^+e^-$ storage ring, if the International Linear Collider is not built~\cite{Bhat}. 

The 100~TeV study should address critical technology and cost feasibility issues and, thus, will provide input for developing a long-term roadmap of high-energy physics. Areas of particular interest to U.S. research groups include beam dynamics, magnets, vacuum systems, and machine protection.

%%%%%%%%%%%%%%%%%%%%%%%%%%%%%%%%%%%%
\begin{figure}[tb]
\begin{center}
\includegraphics[width=0.9\textwidth]{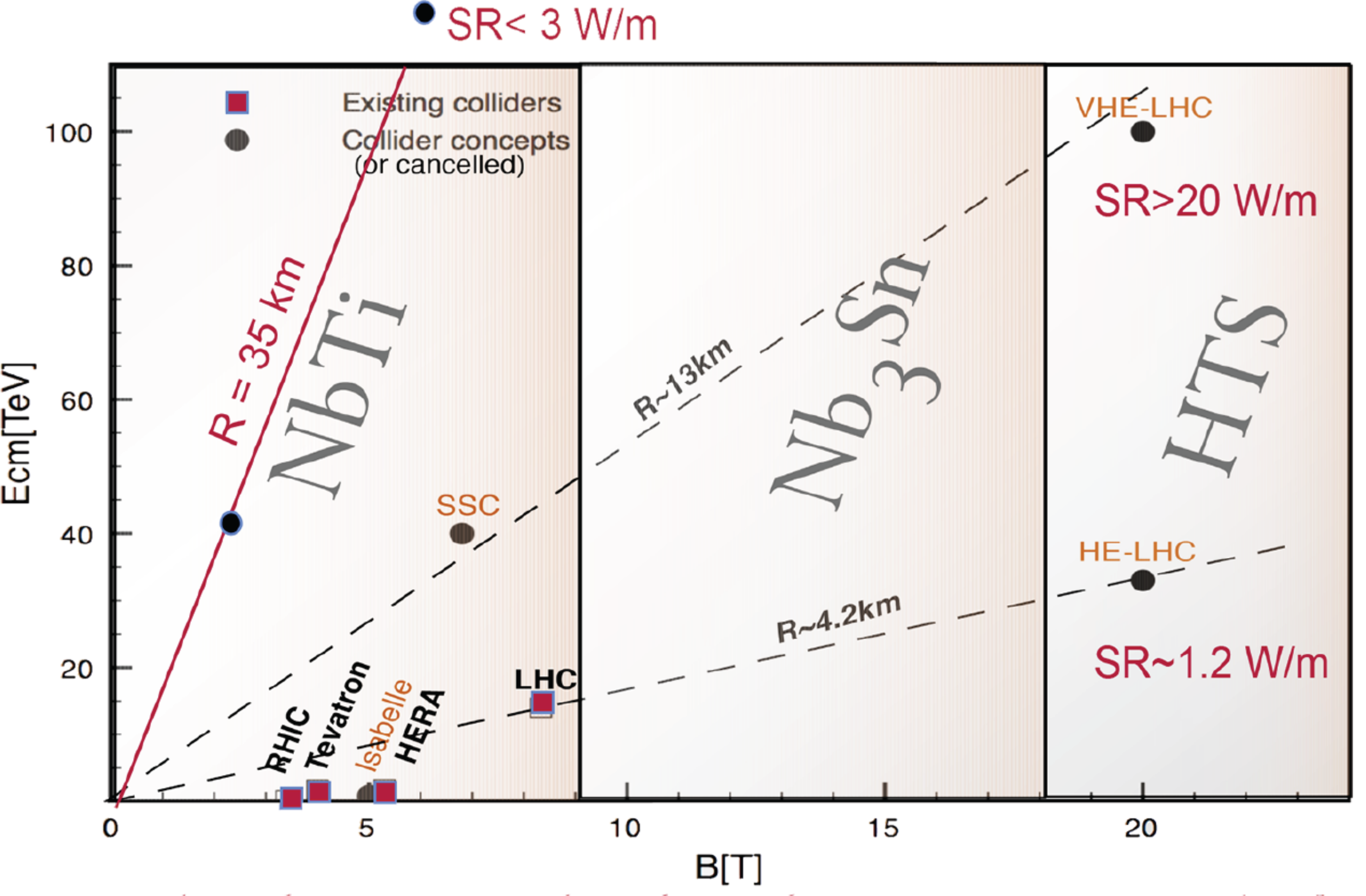}
\caption{Role of the superconductor in the energy reach of hadron colliders as a function of field and bending radius, $R$, produced by the dipole magnets.}
\label{fig:Super}
\end{center}
\end{figure}
%%%%%%%%%%%%%%%%%%%%%%%%%%%%%%%%%

Whether in the LHC tunnel or in a new, larger tunnel a collider with energy beyond the LHC will have to deal with an additional challenge, the emission by the beam of synchrotron radiation at least 20 times greater per meter than the LHC at its nominal parameters. The beam pipe and beam screen will have to absorb that radiation. Although synchrotron radiation is very beneficial for beam stabilization and will make the higher energy collider the first hadron machine dominated by synchrotron radiation damping, the power dissipated in synchrotron radiation must be removed at cryogenic temperature. In LHC it is removed at 5--10 K. One possible solution for the HE-LHC is based on a beam screen at 40--60 K. This approach has no major drawback in principle, although careful design, engineering and prototyping are needed to validate such a design. Merely relying on a beam screen at 10~K similar to the LHC would be a heavy burden for the cryogenics. Handling synchrotron radiation may be even more complex for a 100 TeV machine.

{\it Focused engineering development is no substitute for innovative R\&D. }

In addition to the focused program of engineering development, the separate programs of long-range research in magnet materials and mechanical structures of magnets must be continued. These programs have provided the intellectual and infrastructure base for the success of LARP.  For example, the long-range future of high-energy physics using p-p collisions may need magnet operating fields beyond 16~T.  In practice, the decision will rest on a tradeoff between tunnel cost --- which scales roughly inversely with dipole field --- and the cost of the dipole magnets that scales roughly linearly with dipole field for long magnets.  The crossover in cost cannot be decided ab initio as the tunnel cost will depend strongly on the geology of the collider site. 
A new generation of magnets using high-temperature superconductors will require new engineering materials with small filament size that are available in multi-kilometer continuous pieces.  Advanced magnets may offer greater temperature margin against quenches due to stray radiation lost from the beam. Higher-field magnets will require innovative stress management techniques, exquisitely sensitive magnet protection schemes, and perhaps novel structural materials.

Although synchrotron radiation damping is not dominant in proton colliders, as it is in electron colliders, it is still important to understand beam dynamics and other effects of marginal synchrotron radiation damping. As the energy stored in the beam becomes several gigajoules, control of tenuous beam halos becomes a pressing issue. Likewise machine protection from accidental beam loss and the design of special beam abort dumps become difficult challenges that demand innovation. An effective long-range research and development program must also consider beam physics of the injection chain, effects of noise and ground motion, and design and technology options for the configuration of the interaction regions.

\section{Energy Frontier lepton colliders}

In its first three years of running experiments, the Large Hadron Collider (LHC) has begun to explore the energy region up to 1 TeV. The LHC experiments have discovered a Higgs boson which,  within current experimental precision,  is consistent with that of the Standard Model and have measured its mass as about 126~GeV. The discovery of the Higgs boson and the recent release of a technical design report for the International Linear Collider have energized high energy physicists to evaluate in depth the experiments that could be performed with a lepton collider. As described in the report of the Snowmass Energy Frontier Study Group, a future lepton or photon collider would provide a factory for measurements of the properties of the Higgs boson with ultimate precision.  A lepton collider would allow unambiguous searches for new particles that complement searches at the LHC. It would also provide opportunities to probe for and study new physics, both through the production of new particles predicted by models of physics beyond the Standard Model and through the study of indirect effects of new physics on the $W$ and $Z$ bosons, the top quark, and other systems.

\subsection{International Linear Collider }

In June 2013 the Global Design Effort (GDE) published a Technical Design Report (TDR)~\cite{ILCTDR} of the International Linear Collider (ILC), an accelerator that will give these capabilities. The ILC is a superconducting linear $\ee$  collider with a center-of-mass collision energy tunable between 200 and 500 GeV with a luminosity   exceeding 10$^{34}$\lunit\  at 500 GeV, roughly scaling in proportion to the collision energy. The ILC is upgradable in luminosity by a factor of 2 and upgradable  in energy to 1 TeV.  The Japanese high-energy physics community has named the ILC as its first priority. An experienced cadre of U.S. accelerator physicists and engineers is capable and ready to work on this project. 

{\it Our study welcomes the initiative for ILC in Japan. The U.S. accelerator community is capable of contributing as part of a balanced U.S. high-energy physics program.}

The key characteristics of the ILC accelerator are the relatively long interval between collisions of bunches, narrow beam energy spread, beam position and energy stability, and the ability to polarize both electrons and positrons. The design and technical details of the ILC have been developed over more than 20 years and incorporate extensive US leadership contributions in beam dynamics, damping ring design, electron and positron sources, superconducting RF (SRF) technology,  and beam delivery system design. As described in the TDR, the ILC represents a mature collection of technologies that are ready to proceed to construction.  
The 1.2 GeV VUV Free-Electron-Laser “FLASH” facility at DESY has provided a critical integrated system test experience. FLASH technology is quite similar to ILC, and ILC operational parameters are within the range of FLASH hardware. Tests at FLASH were done with the ILC bunch number, bunch repetition rate, bunch charge and peak beam current. These tests were successful and encountered no fundamental technology issues with operating a superconducting linac with the ILC design characteristics. The research program has successfully demonstrated the goal of 31.5 MV/m in installed cryomodules with beam loading, using niobium cavities with no more than two surface-preparation processing cycles. With the specified accelerating gradient, the total length of the 500~GeV ILC is 31~km.  

Industrial and institutional partners of the GDE advised a strong industrialization program from the outset. A key ingredient was the development, within the GDE, of a project-governance and value-based cost-estimating strategy that was balanced regionally and that took advantage of the intrinsic modularity and relative maturity of the SRF technology. The most costly and time-consuming part of industrialization is the construction and commissioning of heavy infrastructure, notably institutional test facilities. Fabrication of superconducting cavities to the ILC design specifications has been industrialized, with qualified vendors in Europe, North America, and Asia.

%%%%%%%%%%%%%%%%%%%%%%%%%%%%%%%%%%%%
\begin{table}[tb]
\begin{center}
\begin{tabular}{l|l|c|c|c|c|c|c}
$E_{CM}$ & GeV &   250   &  350 & 500 &  500 up &  250 up &  350 up \\ \hline
Rep. rate  &	Hz  &	5  &	5  &	5  &	5  &	10  &	8   \\ \hline
Bunches/pulse   &	 &	1312 &	1312 &	1312 &	2625 &	2625 &	2625\\ \hline
Total beam power &	MW	 & 5.3 &	7.4 &	10.5 &	21.0	 &21.0 &	23.5\\ \hline
Luminosity  &	$10^{34}$\lunit   &	0.75	 & 1.0 &	1.8 &	3.2 &	3.0	 &3.2  \\ \hline
\end{tabular}
\caption{Summary of ILC performance in the range 250 to 500 GeV.   Columns 3--5 give
the TDR values. Columns 6--8  indicate scenarios beyond the parameters discussed in the TDR, but still within its technical scope. Units are specified  in the 
second column.}
\label{tab:ILCparams}
\end{center}
\end{table}
%%%%%%%%%%%%%%%%%%%%%%%%%%%%%%%%%

%%%%%%%%%%%%%%%%%%%%%%%%%%%%%%%%%%%%
\begin{table}[tb]
\begin{center}
\begin{tabular}{l|c|c}
Center-of-mass energy   & \% of TDR cost &	Power consumption (MW) \\ \hline
250 GeV “Higgs factory” &	70\%	 & 120  \\  \hline
500 GeV	& 100\%& 	163 - 204 \\ \hline
1 TeV upgrade &	150\%&	240\\ \hline 
1.5 TeV upgrade &	200\%&	$> 300$\\ \hline
\end{tabular}
\caption{Scaling of ILC cost and power consumption scaling as function of center-of-mass energy.}
\label{tab:ILCscaling}
\end{center}
\end{table}
%%%%%%%%%%%%%%%%%%%%%%%%%%%%%%%%%%%

During the research phase, U.S. national laboratories and industry have contributed substantially to cavity performance, cryomodule design, high-level RF performance, demonstrations of beam dynamics, and specific components and systems. For the construction of ILC, it is reasonable to assume the U.S. contributions will continue along these lines. Specifically, construction and testing of completed cryomodules will contribute the greatest economic value to the project. The US community could also provide detailed accelerator design development efforts as an intellectual contribution. It is expected that the US would also contribute to design and construction of one or more linear collider subsystems. Given that the TDR estimates the required labor effort to be 13,000 person years, participation by America's highly experienced accelerator scientists and engineers will be crucial.

Extension of the ILC to 1 TeV is straightforward. It requires lengthened linac tunnels and additional cryomodules, but would use the original ILC sources, damping rings, final focus and interaction regions and beam dumps. No new technological breakthroughs would be required, although research to develop higher-gradient cavities would permit shorter tunnel extensions and thus cost savings.

\subsection{Alternative $\ee$  colliders}

Alternative approaches to the ILC – albeit on a much longer time scale – include approaches with the potential to reach multi-TeV energies. One approach, the Compact Linear Collider (CLIC)~\cite{CLICCDR}  two-beam accelerator, is based on a high-gradient X-band, warm linac powered by a high current, low energy 
(about 1~GeV) drive beam. CLIC would stretch for 50 km at an ultimate center-of-mass energy of 3 TeV. In contrast, the muon collider --- if proven feasible --- would fit a 3--5~TeV collider onto the present Fermilab site. Moreover, the muon approach has a very large overlap with capabilities needed for Intensity Frontier accelerators. Yet another approach would use wakefields driven either by beams or lasers to achieve accelerating fields of 10 to 100 GeV per meter. Many feasibility and practicality issues need to be solved in order to implement any of these programs.

For CLIC options at 350 GeV, 1.5 TeV and 3 TeV are currently being studied. The first and second option use a single drive-beam generation complex to feed both linacs, while in option three each linac would be fed by a separate complex. The CLIC design is based on three key technologies, which have been addressed experimentally: high gradient structures, drive beam generation, and power extraction. The normal-conducting accelerating structures in the main linac have a gradient of 100 MV/m, to limit the length of the machine. The RF frequency of 12 GHz and detailed parameters of the structure have been derived from an overall cost optimization at 3 TeV. Experiments at KEK, SLAC, and CERN have verified the structure design and established its gradient and breakdown rate. The drive beams run parallel to the colliding beams through a sequence of power extraction and transfer structures, where they produce short, high-power RF pulses that are transferred into the accelerating structures. These drive beams are generated in a central complex. The drive-beam generation and power extraction have been demonstrated in a dedicated test facility at CERN.

Wakefields in plasma-based accelerators can potentially provide a thousand-fold or greater increase in acceleration gradient over standard technologies.  Two primary approaches are being investigated: beam-driven wakefields (plasma wake field acceleration, or PWFA) and laser driven wakefields (laser plasma acceleration, or LPA). Experiments at SLAC, LBNL, and at European laboratories have shown that plasmas can accelerate and focus high-energy beams at an accelerating gradient in excess of 50 GeV/m. At present two large R\&D facilities are spearheading this research in the U.S., FACET at SLAC (PWFA) and BELLA at LBNL (LPA). Smaller efforts are being pursued at university laboratories.

For any variant of wakefield accelerator to be practical as a linear collider, several feasibility and practicality issues must be resolved in the context of an integrated system test. Most importantly, wakefield accelerators, like standard accelerator modules, must be capable of being staged in a series of phase-locked segments.  Both PWFA and LPA approaches must demonstrate simultaneous positron acceleration and focusing in plasma densities consistent with preserving positron beam quality. Both must demonstrate timing, pointing, and focusing control consistent with the high luminosities required in a lepton collider.  Finally, both must demonstrate that multi-bunch plasma instabilities (such as zero-frequency, convective hose instability) can be overcome with operation at the tens of kHz repetition rate required for high luminosity.  Beyond the feasibility issues are questions of practicality related to overall cost, efficiency, and reliability. These issues may be more demanding for laser-driven concepts as cost and efficiency limitations of drive beam sources are better known.

For the study of the Higgs boson what matters are deliverable luminosity, maturity of design, risk, timescale, and cost. The type of $\ee$collider, linear or circular, is of little importance.  A circular collider might offer a potential alternative to the linear collider. It can benefit from three characteristics of lower-energy circular machines: (1) high luminosity and reliability, (2) the availability of several interaction points, (3) excellent beam energy accuracy. For a given RF power, the luminosity of a storage ring collider rises linearly with its circumference. For a given tunnel size, and assuming the machine operates at the beam-beam tune-shift limit, luminosity rises linearly with the total dissipated synchrotron radiation (SR) power, which is approximately proportional to the total available power. Therefore, the analysis can be scaled to different machine sizes. The energy reach consistent with sufficient luminosity for Higgs studies depends strongly on the machine circumference; a machine of 27~km circumference might reach 240~GeV, while the limit is about 
350~GeV for a machine of 80~km. At a given circumference, the energy reach and luminosity are very strongly coupled; detailed characteristics of the design matter greatly.  

CERN has begun a design study of the TLEP storage ring, a machine with a circumference of 80 to 100 km circumference and dissipated synchrotron radiation power set at 100 MW. The luminosity would be largest at the $Z$ peak, but fall rapidly as center-of-mass increases beyond that point.  The proposed TLEP storage ring would need ~ 1 km of superconducting RF cavities and low $\beta^*$ insertions, would operate at fixed field and fed by a synchrotron situated in the same tunnel for continuous top-up injection. Multi-bunch operation would be necessary for high luminosity below the top energy, requiring separated beam pipes for electron and positron beams. Beamstrahlung  from the collision process can lead to a short luminosity-lifetime and large instantaneous energy spread in the beam of the order of several percent. Beam losses must be tightly controlled even in a lattice of very large dynamic aperture.  Maintaining vertical to horizontal emittance ratios considerably smaller than light sources needs further study. 

The Chinese high-energy community has also started discussion and  first studies of a circular Higgs factory or a  proton collider in a very large tunnel~\cite{IHEP}.  Should the ILC not go forward
 over the next decade and should the renewed interest in a very large circumference hadron collider be sustained, the possibility of a circular Higgs factory deserves extensive consideration.

 %%%%%%%%%%%%%%%%%%%%%%%%%%%%%%%%%%%%
\begin{figure}[tb]
\begin{center}
\includegraphics[width=0.9\textwidth]{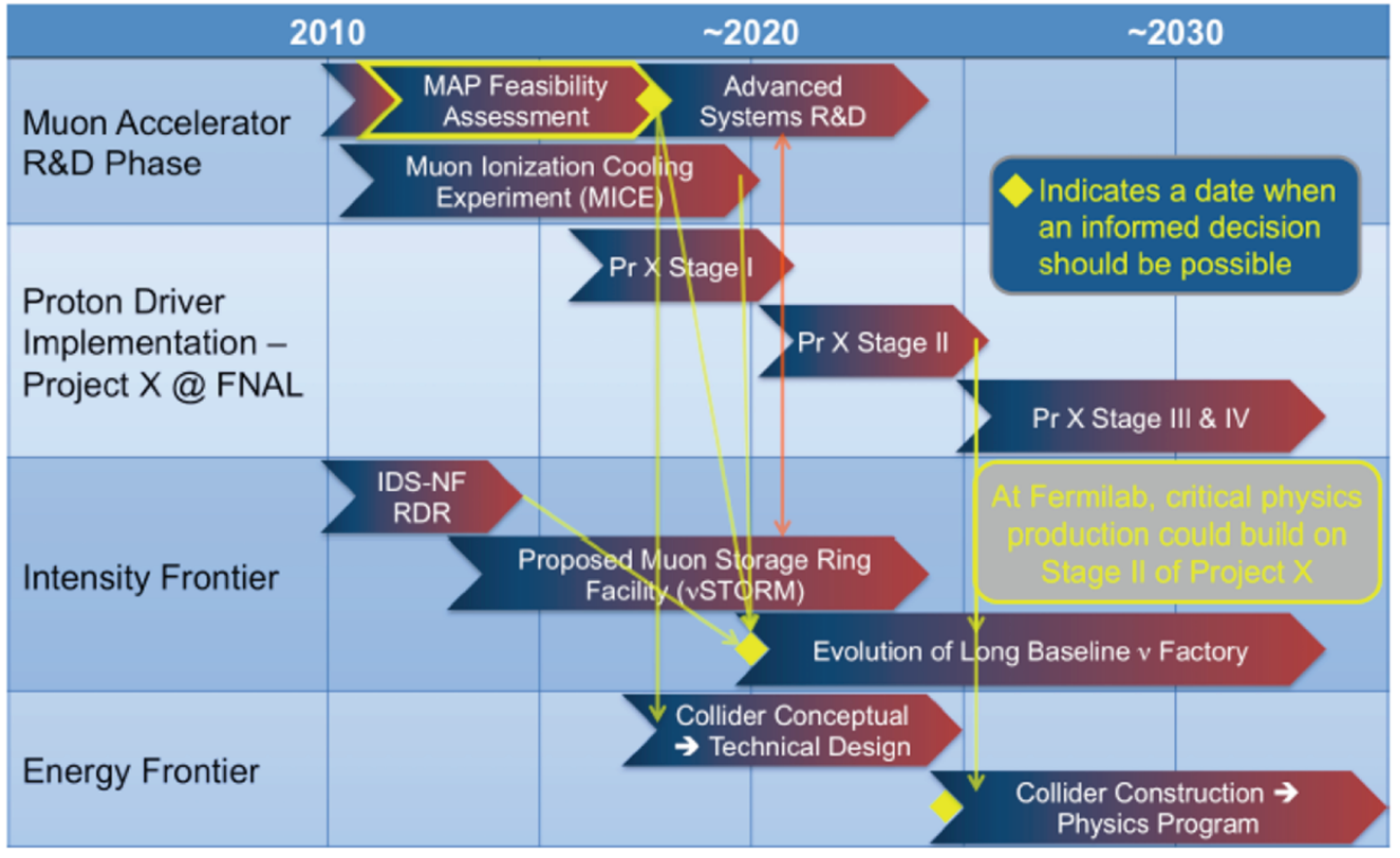}
\caption{ Muon accelerator timeline, including the MAP Feasibility Assessment period. It is anticipated that decision points for moving forward with a Neutrino Factory project supporting Intensity Frontier physics efforts could be reached by the end of this decade, and a decision point for moving forward with a Muon Collider physics effort, supporting a return to the Energy Frontier with a U.S. facility, could be reached by the middle of the next decade.}
\label{fig:Muon}
\end{center}
\end{figure}
%%%%%%%%%%%%%%%%%%%%%%%%%%%%%%%%%

\subsection{Muon accelerators}
Muon accelerators~\cite{Palmer} have the potential to provide world-leading experimental capabilities for physics at center-of-mass energies from the Higgs at 126 GeV up to the multi-TeV scale. A circular muon collider can potentially reach the higher energy range because the larger mass of the muons means they produce far less synchrotron radiation than electrons. For the same reason, beamstrahlung is suppressed leading to a very small spread of the center-of-mass energies in muon collisions. Muons, however, have a lifetime (at rest) of about 2.2 $\mu$s and therefore decay in flight.  The unstable nature of the muon demands that the sequence of beam creation, manipulation, and acceleration be done rapidly; high-gradient acceleration is essential. Therefore, an Energy Frontier muon collider would necessarily be relatively compact. Even a 5~TeV collider would fit within the Fermilab site.

Research of critical importance to the performance of a muon collider includes: (1) development of a high-power target station including a high-field capture solenoid (nominally 20~T hybrid normal and superconducting magnet with about 3 GJ stored energy) that is ultimately capable of handling more than 4 MW of power; (2) control of collective effects in the high-intensity beams at the low energy end of the production, capture, and acceleration process; (3) cooling (reduction) of the muon six-dimensional phase space volume by six orders of magnitude to achieve the beam parameters required for the muon collider designs; (4) development of very large-aperture, superconducting dipole magnets with heavy inserts to protect the superconductors from the decay products of the stored muons in the collider. Once items (1) and (2) are resolved positively, a muon accelerator would provide an outstanding high-flux source of well-characterized neutrinos for a range of Intensity Frontier experiments. The potential for muon accelerators to address crucial questions on both the Intensity and Energy Frontiers argues for a robust development program.

\subsection{Photon colliders}

In a Higgs factory photon collider, two electron beams are accelerated to 80 GeV and converted to about 63 GeV photon beams by colliding with low-energy (3.5 eV photons), high-intensity (5 J per pulse) laser beams via the Inverse Compton Scattering (ICS) process. The two high-energy photon beams then collide and generate Higgs particles through the $s$-channel resonance $\gamma\gamma\to H$. Among various options for a Higgs factory, a photon collider has the distinct advantage that the 80 GeV energy required for the electron beam is lower than for other colliders. Photon colliders have been discussed as options to accompany proposed linear or circular colliders or as a stand-alone facility. 
Since the overall acceleration efficiency of any facility is the product of the individual efficiencies of each of the systems involved in the transfer of power from the wall plug to the beam, each of the systems should be as efficient as possible. Innovative research on efficient RF generation would be extremely beneficial to all lepton collider designs. It should be strongly supported as a key technology to reduce the operating costs of future facilities for both the Energy and Intensity Frontiers.

\section{Capabilities requiring high-intensity proton beams}

The beam parameters needed for a variety of measurements of interest in the Intensity Frontier were presented at a community workshop held at Brookhaven National Laboratory in April, 2013~\cite{Galambos}.
 Experiments included neutrino, kaon, muon, neutron, and proton EDM measurements. From the secondary beam specifications, the workshop identified the proton beam requirements for the different proposed Intensity Frontier measurements. Although the resulting sets of primary proton beam requirements were generally not based on detailed calculations but rather on general guidelines, they can be used as a guide for matching the desired measurements with existing or proposed proton facilities. The study compared 20 existing proton beamlines and 14 planned upgrades with the requirements of 19 secondary beam requests filled out by experiment advocates and drew the following conclusion:

{\it The next generation of Intensity Frontier experiments will require proton beam intensities and timing structures beyond the capabilities of any existing accelerators.}

The major proton accelerator facilities surveyed included U.S. capabilities at FNAL, BNL, LANL, and SNS. International representation included CERN, J-PARC, PSI, ISIS, and TRIUMF. New and proposed proton accelerator facilities included the European Spallation Source (ESS), the Chinese Neutron Spallation Source (CSNS), DAE$\delta$ALUS, and Project X. In addition, existing facilities typically have upgrade proposals --- both near and long term --- which have been included in the study. High proton beam powers of 1 MW or more are desired for most proposed experiments. The beam energy requirement typically exceeds 1 GeV.  The required time structure of the beam varies significantly, depending on the nature of the measurements and the specifics of the measurement conditions; for example, underground detectors have special requirements. 

Accelerators with current and anticipated future programs of Intensity Frontier experiments include FNAL, J-PARC, and CERN. The LANL linac is undergoing a refurbishment and in a few years this beam would operate at 100 Hz and provide bunches with a 1 ms duration available for potential users. Cold neutron applications may be able to use this beam structure, as well as others.  Existing accelerators such as J-PARC and SNS might provide MW proton beams, with energy  less than 3 GeV with very low duty factor (less than 10$^{-4}$ to minimize background from cosmic sources) for lower-energy muon and electron neutrino studies.  ISIS (in the UK) has been used for this purpose in the past, and still has a cavern 10 m from their TS-1 target.  The ultimate technical potential of linac technology is suggested by the example of LEDA. The normal-conducting LEDA linac, which produced a 6.7 MeV, 100 mA CW beam, operated at Los Alamos from 1999 to 2001. 

\subsection{Proposed new accelerators}\label{sec:intenseaccel}

Proposed new accelerators for intensity frontier applications include Project X, DAE$\delta$ALUS, and nuSTORM. Of the three, Project X~\cite{Holmes}  has the broadest scope and, if required, could be structured as a multi-phase proposal. Project X is a high-intensity proton facility that will support a world-leading Intensity Frontier research program over the next few decades at Fermilab. When compared to other facilities in the planning stages elsewhere in the world, Project X would be unique in its ability to deliver, simultaneously, up to 6 MW of site-wide beam power to multiple experiments, at energies ranging from 233 MeV to 120 GeV, and with flexible beam formats. Project X is designed to support a wide range of experiments based on neutrinos, muons, kaons, nucleons, and nuclei. It will also lay the foundation for the long-term development of a neutrino factory and/or a muon collider. 

Project X capitalizes on the very rapid development of superconducting RF technologies over the last 20 years, and their highly successful application to high-power H− acceleration at the Spallation Neutron Source (SNS) at Oak Ridge National Laboratory. As a result of the accelerator research leading to and supporting SNS construction and operation, excellent simulation and modeling tools exist for designing the Project X facility with high confidence that the performance goals can be achieved. The primary supporting technologies required to construct Project X exist today.

The Project X Collaboration is engaged in a comprehensive development program aimed at mitigating technical and cost risks associated with construction and operations. The major elements of the risk reduction program include an integrated, 25 MeV front-end system test, the Project X Injector Experiment (PXIE). To allow for beam accumulation in the Main Injector, the Project X linac will accelerate H$^-$ ions. Optimizing the superconducting RF cavities will improve the cost effectiveness of the final accelerator design. Achieving the performance required of high-power targets is expected to be challenging. Therefore, high-power target technology will be a key component of research in support of the project.

The overall scope and goals of the Project X development program are based on being prepared for a 2017 construction start. Essentially all elements described above are required to implement the first stage of Project X.

DAE$\delta$ALUS~\cite{Conrad}  is a neutrino research program based on ``decay-at-rest'' sources.  Pions are produced by interaction of 800 MeV protons on a suitable target. This energy is sufficiently above threshold for good pion yield, and low enough that pions will stop in the target before decaying.  The DAE$\delta$ALUS configuration consists of three nearly identical
 sources of neutrinos. The highest-power accelerator, located 20 km from the
 detector would provide 10 mA of protons at 800 MeV to the neutrino-generating targets. 
A current of 10 mA is approximately a factor of 5 times the highest achieved current at PSI, the world's leading high-power cyclotron today.  Accelerating H$_2^+$  ions rather than
 protons has the potential for reducing space-charge issues at injection. Extraction of 
the beam at 800 MeV with a stripper foil minimizes the necessity for clean turn separation at the outer radii, only requiring a proton extraction channel with sufficiently large momentum 
acceptance to transport ions stripped from several overlapping turns. The use of H$_2^+$ acceleration represents a novel approach to reducing beam losses at extraction from the
 800 MeV cyclotron.

To date, the DAE$\delta$ALUS feasibility arguments are made by scaling from existing low-energy H$^-$ commercial cyclotrons as well as from the PSI high-power proton cyclotron.  Since the high-power H$_2^+$ concept is quite novel, a systematic research is being conducted to address challenges of meeting the required performance. The most critical elements are (1) ion-source development to achieve very bright, vibrationally cold  H$_2^+$ beams of at least 50 mA CW, (2) injection into cyclotrons with emphasis on bunching efficiency, space requirements and space-charge dynamics, (3) end-to-end simulations to evaluate beam stability and uncontrolled loss, and (4) atomic physics experiments to measure stripping and beam-gas scattering cross sections and possibly techniques for Lorentz dissociation of vibrational states in high-field (about 25~T) magnets in the transport line between the injection and the second-stage cyclotrons. 
As is the case for Project~X, target systems are challenging and will require sustained research.

The first phase of DAE$\delta$ALUS is IsoDAR~\cite{Bungau}, a compact 600 kW cyclotron proposed to be located 15 m from the KamLAND detector for a definitive search for one or two sterile neutrinos.  The IsoDAR cyclotron would also be a prototypical injector for the superconducting ring cyclotron of DAE$\delta$ALUS. At present space charge effects at injection are being studied experimentally in a collaboration with industrial partners.

The nuSTORM facility is designed to deliver beams of electron and muon neutrinos from the decay of a stored muon beam with a central momentum of 3.8 GeV/c and a momentum acceptance of 10\%. The facility will be able to (1) conduct highly sensitive searches for sterile neutrinos, (2) provide definitive measurements of neutrino-nucleon scattering cross sections with percent-level precision for the electron and muon neutrinos, and (3) take the crucial first step in developing of muon accelerators as a powerful new technique for particle physics.

nuSTORM is the simplest implementation of the neutrino factory concept.  It utilizes protons from the Main Injector, has a conventional (NuMI-like) target station, pion collection with a NuMI-style horn, a pion transport line, and a 480 m circumference muon decay ring.  The nuSTORM facility can also provide an intense pulses of $10^{10}$  low-energy muons, appropriate for future 6-dimensional muon ionization cooling experiments.  This beam is available during operation of the neutrino physics program and operates in a completely parasitic mode.  Finally the muon decay ring provides the opportunity for accelerator R\&D on beam instrumentation and potentially on muon acceleration technology.

\subsection{General accelerator research needs}

In addition to project-specific needs, the Snowmass workshop identified more general research needed to advance Intensity Frontier capabilities. That research includes but is not limited to the following~\cite{Shiltsev}:
\begin{itemize}
\item 	Understanding and controlling beam loss: Very precise simulations of beam dynamics involving accurate predictions of beam halos are essential for enabling high-power multi-megawatt-scale machines to run without inducing excessive amounts of radioactivity. To benchmark simulation codes, reliable measurements of very weak beam halos will be required with a large (10$^4$--10$^5$) dynamic range.
\item 	Superconducting RF technology: Most modern high-power proton facilities rely on linacs utilizing superconducting radio frequency (SRF) acceleration. The initial development of SRF technology emphasized maximizing the accelerating gradient to limit the size of the accelerators. Since  many Intensity Frontier applications require the delivery of high-duty-factor beams, the figure of merit is shifting from high-gradient to very high unloaded quality-factor, Q$_0$. Cavity fabrication and processing techniques must minimize cost, while reliably producing cavities with gradients of roughly 20~MV/m at Q$_0$ in excess of 10$^{10}$ to support future facilities.
\item 	High-quality, high-current injection systems: Multi-MW linacs and cyclotrons will need high-current, low emittance sources of ion beams, effective beam chopping, and control of strong space-charge effects at low beam energies. While suitable ion sources are close to being available, effective beam chopping is beyond the present state of technology and is under active development. The challenge is to develop choppers to provide full beam deflection on a time scale less than the beam micro-bunch spacing, and/or to cleanly transport partially deflected beam.
\item 	Isochronous ring cyclotrons: Isochronous ring cyclotrons (including fixed-field alternating gradient accelerators) are also good candidates for high continuous (CW) power. Clean extraction of the beam requires well-separated turns. This sets the scale of both the ring diameter and the energy gain per turn. The alternative is to use an ion such as H$_2^+$  that can be extracted with a foil. The large superconducting magnet sectors require some development beyond the present RIKEN magnets before costs can be confirmed. 
\end{itemize} 

\subsection{High-power targets}

High-power target development and its implications for high-power proton accelerators should not be underestimated~\cite{Hurh}. Three of the present high-power accelerator facilities (SNS, J-PARC-MLF, and FNAL-MINOS) have been limited in beam power during recent operational periods by target concerns, not by the accelerator capability. The cost of these systems is a non-negligible portion of high-power accelerator facilities, and the resources needed to operate and maintain them are also significant (remote maintenance, disposal of radioactive wastes, control of personnel exposure to radiation). Plans for new facilities should include the impacts of handling the proposed high beam powers.

Of the many identified challenges of high-power targets, radiation damage is of special note because it is complex, difficult to evaluate, and time-consuming to investigate. Atom displacements and gas production are the main underlying damage mechanisms, the particulars of which depend on primary beam characteristics, target materials, and specifics of the mechanical design. In the regimes of beam energy and intensity foreseen for Intensity Frontier experiments, one cannot simply scale from experience with nuclear power reactors. Moreover, the formation and growth of defects in target materials in extreme conditions is a mesoscale physics problem that is presently beyond present modeling and computational capabilities.  In addition, techniques of accelerated testing are frequently not validated and/or appropriate irradiation facilities may not be available.

{\it Sustained, focused research into target technology in the context of a broad international collaboration of interested laboratories will be essential.}

Experiments on the Intensity Frontier offer a broad range of scientific opportunities accompanied by a similarly broad range of technical challenges that must be overcome to conduct these experiments in acceptable time frames. The technical challenges will require sustained, focused efforts. The time to begin is now.

\section{High-intensity electron and photon beams}

To offer new opportunities for physics discoveries heavy flavor factories must accelerate and store beams with characteristics that will be challenging to deliver. Two types of flavor factories are being constructed or proposed worldwide: super $B$ factories and super tau-charm factories. SuperKEKB, a super-high-luminosity $B$ factory, is an upgrade to the KEKB $B$-factory currently under construction in Japan, with commissioning scheduled to commence in January 2015. 
U.S. national labs and universities have made important contributions to the design. Two super tau-charm factories have been proposed: one at Frascati (Tor Vergata) in Italy and one at Novosibirsk in Russia. Both machines are two-ring, symmetric-energy machines, with provisions for longitudinally polarized beams.

For SuperKEKB to achieve its target luminosity of $8\times 10^{35}$\lunit\ --- a 40-fold increase over that of KEKB  --- it will require  beam currents approximately twice as high as used at KEKB and vertical bunch sizes at the collision point about 20 times smaller than those achieved at KEKB. Although it would be premature to contemplate increases in the design luminosity of a machine that has not yet started commissioning, there is performance headroom in machine subsystems for increasing luminosity beyond the its baseline 
design value~\cite{Groupfour}. Continuing U.S. collaboration would strengthen the SuperKEKB project, and would support efforts to increase the luminosity to well above that targeted in the existing design. This result would mirror the operational experience with both KEKB and PEP-II.

Enhanced collaboration could also enable polarized beams in the collider. Polarization in SuperKEKB will involve preparing and injecting a beam of polarized electrons in the correct spin orientation parallel to the local precession axis at the injection point. While polarized sources with the required intensity exist (e.g., the SLC polarized gun), they are not low-emittance polarized sources. BNL has developed a candidate gun which is ready for commissioning and testing. The alternative would be to use a polarized DC gun followed by a damping ring to lower the emittance. The physics case needs to be made for both higher luminosity and polarized beams.  

The technology development for super flavor factories exploits strong synergies with light sources and damping rings. Relevant areas of synergy include high top-up rate to compensate for low lifetimes, beam stability and control, control of electron cloud and fast ion instabilities, coherent synchrotron radiation issues with short bunches, timing jitter and the attendant energy jitter, and beam instrumentation. Many of these areas are important to the design of damping rings for a linear collider and would certainly apply to the design of a circular Higgs factory. Participation of U.S. accelerator personnel in commissioning and machine studies would help to maintain a high level of U.S. expertise with high-current, low-emittance electron storage rings, such as TLEP.

Three plans or proposals address physics opportunities using high-power electron and positron beams. The DarkLight experiment plans to use the 100 MeV, 10 mA electron beam of the Jefferson Lab free electron laser (FEL), impinging on a hydrogen target, to search for gauge bosons associated with Dark Force theories. The gauge bosons are predicted to decay into $\ee$, leading to the final state $e^- + p + e^+ + e^-$. DarkLight will be unique in its ability to detect all four particles in the final state. The leptons will be measured in a large high-rate time-projection chamber (TPC)
 and a silicon detector will measure the protons. A 0.5~T solenoidal magnetic field provides the momentum resolution and focuses the copious M\o ller scattering background down the beam line, away from the detectors. A first beam test has shown the FEL beam is compatible with the target design and that the hall backgrounds are manageable. The DarkLight experiment has been approved by Jefferson Lab for first running in 2017.
High-intensity electron beams may also be instrumental in fixed target searches for light, weakly-coupled particles including dark matter in the MeV to GeV range. One proposed experiment would require 10$^{20}$ to 10$^{23}$ electrons with energies of $\geq$1 GeV striking stationary nuclei in a beam dump to produce new weakly interacting states, which pass through shielding material and scatter tens of meters downstream in a relatively small detector, one cubic meter or less.
A more speculative proposal would use a high-intensity, low-emittance positron beam, impinging on a plasma target, to generate muon/anti-muon pairs. The beam energy should be just above the $\mu^+\mu^-$ production threshold with maximal beam energy asymmetry, i.e., a $\sim$45 GeV positron beam interacting on an electron target. A low-power experimental test of this muon production rate may be possible at the FACET facility at SLAC.

\section{Electron-ion colliders}

Several designs of future electron-ion colliders have been under consideration in recent years. All of them are based at an existing accelerator facility. The collider configurations include both ring-ring and linac-ring options. Center-of-mass energies of the proposed colliders range from 14 GeV to 2000 GeV.  Most of the collider concepts share enabling technologies. In all designs superconducting RF cavities must be able to operate with high average and high peak beam currents, providing effective damping of high-order modes. The cryomodule design must be able to accommodate containing high beam power. The hadron beam transverse and longitudinal emittances must be small to achieve a high electron-ion collider luminosity. In the designs with medium hadron energy, this requirement calls for the application of powerful cooling techniques.
The low β∗ interaction region designs for all proposed colliders face the issues of strong focusing of beams at the collision point and fast separation of beams after the collision. The synchrotron radiation fan produced by electrons in the interaction region magnets must be kept from hitting the beam pipes whether inside or in the vicinity of the detectors or in superconducting magnets.  Requirements imposed by the detector integration must be satisfied and chromatic corrections to the beam optics have to be implemented while maintaining an acceptable dynamic aperture.  Approaches considered include the following:
\begin{itemize}
\item	Special design of interaction region magnets, including Nb$_3$Sn superconductor technology.
\item	Collimators, absorbers, and masks at appropriate locations to provide protection from synchrotron radiation.
\item	Integrated dipole field in the detector design.
\item	Modified crossing angle geometry (in some designs), including crab-crossings.
\end{itemize}
The linac-ring designs utilize a high-current polarized electron source, with the average current ranging from 6~mA to 50~mA. A Gatling gun arrangement or large-size cathode gun can likely produce up to 50~mA current. The linac-ring scheme introduces non-standard beam-beam effects, which must be explored to understand the limits on the luminosity and the beam parameters. The effects include the electron beam disruption, the hadron beam kink instability, and the effect of the electron beam parameter fluctuation on the hadron beam.
Other shared technologies include polarized beam sources and techniques to preserve beam polarization. In ring-ring colliders spin matching and the harmonic correction techniques have to be investigated further with the goal of minimizing the beam depolarization due to synchrotron radiation, especially in the presence of spin rotators and solenoidal detector magnets.

\section{Accelerator R\&D and test beam capabilities}

This study identified a broad range of test capabilities present in existing or needed for proposed Energy and Intensity Frontier accelerators. The first category of test facilities permits testing beam physics or accelerator components essential to manage technical risks in planned projects. A second category integrates accelerator systems to provide proof-of-practicality tests.  The third category provides tests of physics feasibility of concepts and/or components.  The study identified 35 existing test facilities~\cite{Groupsix} in the U.S. and overseas both with and without beam testing capability. Although these facilities provide substantial readiness to move forward with the highest priority accelerators for high-energy physics, the long-range future of high energy physics needs a few additional dedicated test capabilities in the near term.

\subsection{Energy Frontier accelerators}

Hadron colliders have dominated the high-energy physics landscape for several decades and are expected to continue in that status over many more years of the LHC operation. Several Energy Frontier hadron colliders are under consideration now, including upgrade of LHC luminosity to $5\times 10^{34}$ \lunit, upgrade of the LHC energy to more than 25 TeV, and an 80 to 100 TeV VLHC in a 80--100 km or longer tunnel.
The most crucial technology issues for hadron colliders are in the area of superconducting magnets which are the largest drivers of the cost and of the energy and luminosity reach of these colliders. Of greatest challenge are Nb$_3$Sn 15~T arc dipoles and possible HTS superconductors such as YBCO and Bi-2212 capable of supporting dipole fields well above 20 T. For the VLHC, cheaper lower-field magnets and innovative low-cost tunneling technologies are of utmost importance to establish the financial practicality of a collider. Superconducting crab-cavities and very high-power beam dumps need to be developed for beams of such high energy. 

Many engineering challenges will be associated with the expected few GJ of beam energy in colliders beyond the present LHC. Increased levels of synchrotron radiation will affect beam dynamics, machine design, and operating costs and complexity. These issues will require research to identify methods of efficient collimation and control, novel vacuum chamber designs, and development of the photon stops to protect the superconducting magnets and other equipment.  

Many requisite technology developments for Energy Frontier hadron colliders – described previously  – can be carried out at the existing superconducting magnet and RF test facilities at the DOE national laboratories, as well as at similar facilities overseas, foremost of which is at CERN. Some of these facilities may need upgrades to accommodate the research needs of specific projects (cable tests, transmission line magnet tests, etc.). Many beam dynamics studies concerning collimation and coherent instabilities can be performed at the LHC and its injectors, at the Fermilab Main Injector and Booster, and at RHIC. Most Project X research and development activities and facilities will directly address many issues related to the injector chain for a hadron collider beyond the LHC.  Several promising, novel accelerator schemes – such as resonance-free, integrable optics, space charge compensation with electron devices, and optical stochastic cooling – require a dedicated test accelerator for evaluation. The Integrable Optics Test Accelerator (IOTA) at the Advanced Superconducting Test Accelerator (ASTA) facility at Fermilab is a suitable match for the task. 

Although a primary challenge of any Energy-Frontier $\ee$ linear collider project is value engineering to reduce costs, specific technical challenges remain for design of the positron source and the beam delivery system. These issues are being addressed at facilities including ATF2 at KEK, FLASH at DESY, and ASTA at Fermilab. A CLIC-type system has additional challenges to develop and industrialize the accelerating structures and major components. These questions are being actively addressed at CesrTA, ATF2 and CTF3.

At energies less than 250 GeV per beam, in tunnels with sufficiently large circumference, circular colliders may potentially provide a high luminosity while drawing on a base of mature technology with rich experience. Primary challenges for rings are the large synchrotron power, beamstrahlung, and large momentum acceptance required. Many of these issues are being addressed in large part at flavor factories, diffraction-limited light sources, and beam test facilities.

Photon collider designs allow access to the CP property of the Higgs boson, and they can study it at a lower electron beam energy with a high degree of polarization without the need for positrons and ancillary systems. Primary challenges for photon colliders relate to generating sufficient photon flux with high-energy lasers as well as new interaction region design. Appropriate laser technology is being developed within the international consortiums and at LLNL in the U.S.

A muon collider scales well to high energies because the relatively large mass of the muons allows for tight bends and compact accelerator geometry. The primary challenge for muons is creating and cooling them to sufficient phase space density to allow for high enough luminosity. The Muon Accelerator Program is coordinating the R\&D into the major issues with experiments primarily at FNAL and RAL.

The main challenges for plasma accelerators include demonstrating a concept for positron acceleration and demonstrating staged acceleration with the required beam brightness and efficiency, and those are being addressed through multi-institution collaborations conducting experiments at dedicated and proposed facilities BELLA/BELLA-II at LBNL and FACET/FACET-II at SLAC. 

A major challenge in applying plasma wakefield accelerator concepts to high-energy colliders is staging successive drive beams to accelerate a sufficiently intense witness beam to the desired energy. Staging requires the generation of high-intensity, multiple-drive electron/laser beams capable of high-repetition-rate operation. Tailoring the longitudinal characteristics of a drive beam may effectively increase the energy transfer efficiency from the drive beam to the witness beam, thus reducing the number of required stages. Issues of wakefield acceleration in dielectric structures include identifying materials with appropriate breakdown voltage characteristics, identifying optimal structure geometries, and demonstrating staging with the required efficiency. Relevant research is being conducted at the AWA at ANL, ATF at BNL and FACET at SLAC. The NLCTA facility at SLAC hosts a dedicated beamline studying the primary issues for laser-driven microstructures including damage limits, coupling geometry, staging, and maintaining phase tolerances.

Several facilities in the U.S. are exploring wakefield accelerators. AWA at Argonne combines multiple 75 MeV drive beams with a 15 MeV witness beam. BELLA at LBL uses a 1 PW laser to conduct experiments that synchronize two laser beamlines to accelerate a single, sub-GeV electron beam. FACET at SLAC uses a 30~GeV beam from the SLAC linac to investigate positron acceleration, high-transformer ratio configurations and the ``afterburner'' concept. Multiple drive beam generation and pulse shaping is one of the major goals for the proposed FACET II upgrade.

\subsection{Intensity Frontier accelerators}

The short- and medium-term needs of the U.S. Intensity Frontier program will employ Fermilab proton accelerators --- power-upgraded Booster and Main Injector augmented
 by the ``muon campus'' rings ($g-2$, mu2e, nuSTORM, etc.). Longer-term aspirations require construction of the proposed Project X, described in Section~\ref{sec:intenseaccel}.
 Project X could support a wide range of experiments based on neutrinos, muons, kaons, 
nucleons, and nuclei and also can form a basis of the front-end of a future Neutrino
 Factory. The proposed DAE$\delta$ALUS experiment calls for construction of several 
very high-power, 1 to 5~MW, 
800~MeV proton cyclotrons.    
The primary design constraints in high-power hadron accelerators for the Intensity Frontier are minimizing beam loss ($<$ 1 watt/m, or $<$ 0.01\% total beam loss), and flexible beam delivery in experiment-specific time formats ranging from quasi-CW for rare particle decays to single few-ns-long bunches. Additional challenges facing designers of Intensity Frontier accelerators include: (1) generating high-current, high-quality beams, (2) efficiently accelerating high beam currents to high energies, (3) controlling emittance and minimizing beam halo, (4) manipulating beams on a bunch-by-bunch basis for parallel experiments, (5) maintaining low loss during beam extraction, and (6) creating target systems for extreme power densities and extreme radiation environments.

Project X R\&D activities at the PXIE facility and beam studies with the SRF cryomodules at Fermilab's ASTA facility will directly address many issues listed above. Many beam experiments on efficient collimation, coherent instabilities, and novel extraction methods can be performed at Fermilab's Main Injector and Booster, at the ORNL's SNS, or at the LHC and its injectors. Studies of promising novel accelerator techniques --- like halo suppression in the integrable optics and space charge compensation with electron devices ---
 can be performed at the IOTA at the ASTA facility. A dedicated facility is needed for development and tests of high-power targets, though some target-related issues can be explored at existing accelerators. Many tests and technology demonstrations to support the DAE$\delta$ALUS experiment design work can be performed at the existing facilities, such as INFN-Catania, RIKEN, ORNL, or in the private sector at medical cyclotron companies.

\section{Summary remarks}

The approved LHC program, its future upgrade towards higher luminosities (HL-LHC), and the study of an LHC energy upgrade or of a new proton collider delivering collisions at center-of-mass energies up to 100 TeV, are all essential components of exploring the Energy Frontier with proton colliders. Therefore the priority recommendation of our study is full exploitation of LHC.  Doing so requires a strong LHC Accelerator Research Program sponsored by the OHEP that transitions to a US-LHC high luminosity construction project. During the project period we recommend continuing a focused, integrated, laboratory program that emphasizes the engineering readiness of technologies suitable for a 26 TeV upgrade of the LHC or a machine of higher energy in a larger tunnel. The most critical technology development toward higher-energy hadron colliders is the next generation high-field Nb$_3$Sn magnets (limited to about 15 T) and adequate beam control technology to assure machine protection.

Our study welcomes the initiative for ILC in Japan. The U.S. accelerator community is capable of contributing  as part of a balanced U.S. high-energy physics program. As described in its Technical Design Report (TDR), the ILC is technically ready to proceed to construction. The TDR incorporates leadership U.S. contributions to machine physics and technology in superconducting RF (SRF), high-power targets for positron production, beam delivery, damping ring design, and beam dynamics such as electron cloud effects. 
Plasma-based accelerators (beam or laser driven) providing GV/m accelerating fields would open the way to linear colliders in the muli-TeV range with a relatively small footprint if the preservation of beam quality during multi-stage acceleration can be demonstrated. A circular muon collider, if feasible, could provide discovery-level experimental capabilities at energies from the Higgs $s$-channel threshold at 126 GeV up to the multi-TeV scale. Such a collider could probe narrow physics features with extreme precision, because the larger mass of the muon suppresses synchrotron radiation during the collision process that would spread the energy of the beams. 

Required accelerator capabilities for Intensity Frontier  experiments are more diverse than for the energy frontier. The common characteristics required are (1) average beam power $>$ 1 MW delivered 
with (2) a flexible, experiment-dependent time structure. The overarching conclusion of the study is that the next generation of intensity frontier experiments requires beam intensities and timing structures beyond the capabilities of any existing accelerator. Fermilab's proposed, multi-stage Project X would yield a world-leading facility that would simultaneously deliver a flexible ``on-demand'' beam structure that could serve multiple experiments over an energy range 0.25 -- 120 GeV. Our study heard exciting possibilities for
 DAE$\delta$ALUS, a capability of narrower experimental scope than Project X. For both proposed 
capabilities high-power targets are a difficult challenge that limits facility performance. One cannot 
simply scale the performance with high-energy beams from experience with nuclear reactors. 
Computed radiation effects in inhomogeneous materials, subject to time-varying irradiation, 
need validation with controlled, instrumented in-beam tests. 

The relevant technologies of super flavor-factories exploit strong synergy with light sources and damping rings for lepton colliders. Doubtless the operation of such machines such as SuperKEKB, the super-high-luminosity $B$ factory,  will inform the final design of damping rings for a energy frontier lepton collider. 
All electron-ion colliders studied in recent years would be based at an existing accelerator facility. The collider configurations include both ring-ring and linac-ring options; center-of-momentum energies 
range from 14~GeV to 2000~GeV.  Most of the collider concepts share enabling technologies. 
Superconducting RF cavities must be able to operate with high average and high peak beam 
currents, providing effective damping of high-order modes. 

Although existing accelerator test-bed facilities provide substantial readiness to advance the 
highest priority accelerators for high-energy physics, the long range future of high energy physics 
needs a few additional dedicated test capabilities in the near term, most notably, for high power 
targetry, muon accelerators, and high-intensity accelerator systems.  The group did not examine 
the process in basic accelerator physics and modeling that must accompany any long-range research 
program. Such research bears fruit beyond the confines of specific projects and is essential to the
 long-term health of the field.

The Snowmass Accelerator Capabilities Study offers the following summary findings and recommendations:
\begin{itemize}
\item 	To maximize the exploration of the Energy Frontier, the full exploitation of the LHC is the highest priority of the hadron-collider program. 
\item	As described in its Technical Design Report (TDR), the ILC is technically ready to proceed to construction.
\item 	A vigorous, integrated U.S. research program toward demonstrating feasibility of a muon collider is highly desirable. The current funding level is inadequate to assure timely progress.
\item 	The overarching conclusion is that the next generation of Intensity Frontier experiments requires beam intensities and timing structures beyond the capabilities of any existing accelerator.
\item 	Fermilab's proposed, multi-stage Project X would yield a world-leading facility that could serve multiple experiments over an energy range 0.25 -- 120 GeV.
\item 	Sustained, focused research into high-power target technology in the context of a broad international collaboration of interested laboratories will be essential to frontier accelerator facilities.
\item 	The relevant technologies of super flavor-factories exploit strong synergy with light sources and damping rings for lepton colliders.
\item 	All electron-ion colliders studied in recent years would be based at an existing accelerator facility with center-of-momentum energies range from 14 GeV to 2000 GeV.
\item 	Innovations in acceleration and beam transport techniques such as plasma and dielectric wakefield acceleration have a significant potential to reduce the size of future accelerator facilities. 
\item 	Long-term research including fundamental accelerator and beam physics theory and simulation will expand the technical options for any future accelerator-based facility. Focused engineering development is no substitute for innovative R\&D. 
\end{itemize}

{\bf Acknowledgements: }
This report has been partially supported by the Office of High Energy Physics and the Office of Nuclear Physics of the U.S. Department of Energy and the National Science Foundation. The conveners thank all those who have participated in the Snowmass 2013 process, both at workshops and by submitting white papers.

%\input Theory/snowmass_theory_panel_report_final.tex

%\input ChargedLeptons/wgreport.tex
%\input HeavyPhotons/wgreport.tex

%%%%%%%%%%%%%%%%%%%%%%%%%%%%%%%%%%%%%%%%%%%%%%%%%%
%%%%%%%%%%%%%%%%%%%%%%%%%%%%%%%%%%%%%%%%%%%%%%%%%%
%%%   Your subdirectory (here Magnetism) should include
%%%    the files:
%%%           wgreport.tex
%%%           authorlist.tex
%%%         and all needed figures in pdf format
%%%%%%%%%%%%%%%%%%%%%%%%%%%%%%%%%%%%%%%%%%%%%%%%%%%%
%%%%%%%%%%%%%%%%%%%%%%%%%%%%%%%%%%%%%%%%%%%%%%%%%%%%

\end{document}